\documentclass[prl,showpacs,twocolumn]{revtex4}
\usepackage{amsmath}
\usepackage{graphicx}
\usepackage{amsfonts}
\usepackage{amssymb}

\begin{document}

\title{Morphology Effectively Controls Singlet-Triplet Exciton Relaxation and Charge Transport in Organic Semiconductors}
\author{V.~K.~Thorsm{\o }lle$^{1,2}$, R.~D.~Averitt$^{1,3}$, J.~Demsar$^{1,4,5} $,
D.~L.~Smith$^{1}$, S. Tretiak, R.~L.~Martin$^{1}$, X.~Chi$^{1,6}$,
B. K. Crone$^{1}$, A.~P.~Ramirez$^{1,7}$ and~A.~J.~Taylor$^{1}$}
\affiliation{$^{1}$Los Alamos National Laboratory, Los Alamos, NM
87545, USA} \affiliation{$^{2}$Ecole Polytechnique
F$\acute{e}$d$\acute{e}$rale de Lausanne, CH-1015 Lausanne,
Switzerland} \affiliation{$^{3}$Boston University, Boston, MA 02215,
USA} \affiliation{$^{4}$ Dept. of Physics and CAP, Universit\"{a}t
Konstanz, D-78457 Konstanz, Germany} \affiliation{$^{5}$ Dept. of
Physics and CAP, Universit\"{a}t Konstanz, D-78457 Konstanz,
Germany}\affiliation{$^{6}$Dept. of Complex Matter, Jozef Stefan
Institute, SI-1000 Ljubljana, Slovenia} \affiliation{$^{7}$Bell
Laboratories, Alcatel-Lucent, 600 Mountain Ave., Murray Hill, NJ
07974, USA.}
\date{\today}

\begin{abstract}
We present a comparative study of ultrafast photo-conversion
dynamics in tetracene (Tc) and pentacene (Pc) single crystals and Pc
films using optical pump-probe spectroscopy.  Photo-induced
absorption in Tc and Pc crystals is activated and
temperature-independent respectively, demonstrating dominant
singlet-triplet exciton fission. In Pc films (as well as
C$_{60}$-doped films) this decay channel is suppressed by electron
trapping. These results demonstrate the central role of
crystallinity and purity in photogeneration processes and will
constrain the design of future photovoltaic devices.
\end{abstract}

 \maketitle




\vspace*{-10pt} The strong potential for new technological
applications provided by organic semiconductors has spurred
extensive research efforts in these materials
\cite{ElectProc1999,Hegm2003,GershReview2006}. Applications include
thin film transistors \cite{Forr2004}, light-emitting diodes (OLEDs)
\cite{LightEmit1990,Godl2007}, photodiodes and photovoltaics (OPVs)
\cite{Graetzel1994,Service2004}. To successfully utilize organic
semiconductors in these technologies it is important to understand
both the nature of the photogenerated states and their relaxation
dynamics. In particular, for OPVs it is vital to  understand and
control exciton lifetimes at the materials level in order to fully
optimize efficiency. Such studies involve understanding the role of
crystallinity, defects, and molecular constituency in controlling
relaxation dynamics in the time domain of relevance for
photoconversion \cite{Forr2007,Shao2004}. Here, we present a
comparative study of ultrafast photogenerated state dynamics in
pentacene (Pc) and tetracene (Tc) single crystals and in pure and
C$_{60}$-doped Pc films using optical pump-probe spectroscopy. We
demonstrate for the first time how morphology may control relaxation
behavior in an organic semiconductor by effectively turning
singlet-triplet exciton fission decay channels on and off.
Importantly, we show that triplet suppression, which is necessary
for free carrier formation and thus photovoltaic and photodiode
performance, is controllable via sample morphology.

Exposure of an organic semiconductor to light above the absorption
edge gives rise to various photoexcited species. There is an ongoing
debate in the literature as to the nature of these photoexcitations
and their relaxation dynamics. This includes details of exciton
formation versus the generation of charged polarons or free charge
carriers \cite{Ostro2005,VKT2004,Hendry2004}. These processes are
important in determining the behavior of many organic
semiconductor-based devices such as solar cells and photodiodes.
Organic polymers, such as poly($p$-phenylene-vinylene), have been
extensively studied providing observations of polarons, interchain
excitons (indirect excitons or bound polaron pairs), self-trapped
excitons, triplet excitons and charge transfer dynamics
\cite{Scholes2006,Dex2000,Frolov1997,Bredas2004,Yu1995}. In contrast
to organic polymers with long chains, organic molecular crystals
consist of shorter molecular units and form well-defined crystal
structures. The intermolecular interactions are weak, and the
excitons are largely confined to single molecules resulting in large
exciton binding energies \cite{Claudia2005}. Polyacene organic
crystals such as tetracene (Tc) and pentacne (Pc) are model systems
for studying the intrinsic properties of exciton dynamics. Tc and Pc
molecules consist of, respectively, 4 and 5 benzene rings fused
along their sides and arranged in a herringbone stacking arrangement
with two molecules in each unit cell \cite{Hegm2003}. Tc has an
orange color and luminesces strongly when illuminated due to prompt
and delayed fluorescence \cite{Tomki1971}, while Pc is opaque and is
nonluminescent \cite{Jundt1995}. In the polyacene series of organic
crystals, the energy level of the lowest triplet exciton
$E(\mbox{T}_{1})$ decreases faster than the lowest singlet exciton
energy $E(\mbox{S}_{1})$ with
increasing molecular size. The energy difference $E(\mbox{S}_{1}%
)-2E(\mbox{T}_{1})$ is $-1.3$~eV in napthalene (Nph), $-0.55$~eV in
anthracene (Ac), $-0.21$~eV in Tc, and $0.11$~eV in Pc
\cite{Jundt1995}. In Pc, the excitonic fission process from the
lowest singlet exciton to a pair of the lowest triplet excitons
$\mbox{S}_{1}\rightarrow2\mbox{T}_{1}$ is energetically allowed,
while in Tc this same process is only possible by thermal
activation. This process is strongly suppressed in Nph and Ac. An
energy level diagram for Tc and Pc is presented in Fig. 1. Direct
fission from higher-lying singlet states,
$\mbox{S}_{N}\rightarrow2\mbox{T}_{1}$, is another probable
relaxation channel which competes with nonradiative relaxation to
the lowest excited state, $\mbox{S}_{1}$ \cite{ElectProc1999}.
Charge transfer processes, resulting in carrier generation, compete
with both singlet and triplet exciton formation. In some device
applications, such as solar cells and photodiodes it is beneficial
to enhance carrier generation.

In addressing the importance of these processes we find
singlet-triplet fission to be the dominant process in both Pc and Tc
crystals with a prominent long-lived photoinduced absorption (PIA)
peak originating from $\mbox{S}_{1}\rightarrow2\mbox{T}_{1}$ and
$\mbox{S}_{N}\rightarrow 2\mbox{T}_{1}$ fission. In contrast to Pc,
the triplet production in Tc is strongly temperature dependent in
agreement with the thermally activated
$\mbox{S}_{1}\rightarrow2\mbox{T}_{1}$ fission process, where
fission competes with radiative fluorecence. In comparison to Pc
crystals, we find that in Pc films the triplet production is
quenched and the dynamics is largely dominated by charge transfer
originating from defects (electron acceptors). This was further
supported by measurements on C$_{60}$-doped Pc films, with C$_{60}$
being a known electron acceptor \cite{Yu1995}.

\begin{figure}[ptb]
\centering
\includegraphics[width=1.0\columnwidth]{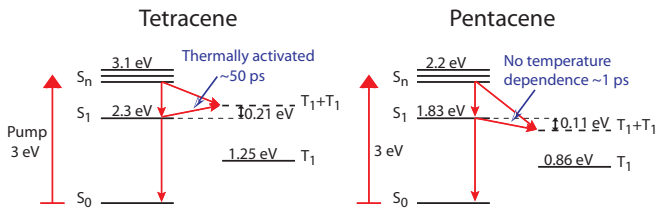}
\caption{Energy level diagram of Tc and Pc. Only the levels
participating in the singlet-triplet fission processes following
photoexcitation at 3.0 eV are shown. Here, $\mbox{S}_{N}$ represents
higher-lying states in the singlet manifold, $\mbox{S}_{1}$ the
lowest singlet exciton, and $\mbox{T}_{1}$ the lowest triplet
exciton. $\mbox{T}_{1}+\mbox{T}_{1}$ denotes two triplet excitons
produced by fission. Important relaxation channels are denoted by arrows.\vspace*{-10pt} }%
\label{Fig: 1}%
\end{figure}

\begin{figure}[ptb]
\centering
\includegraphics[width=0.8\columnwidth]{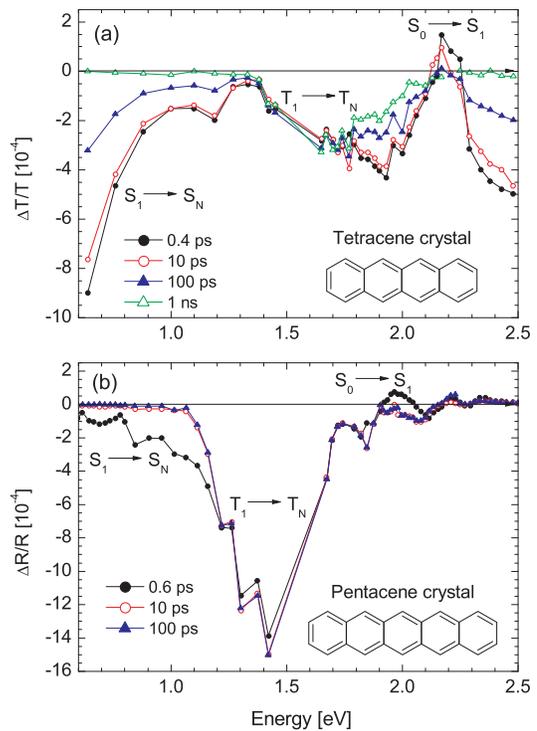}
\caption{Transient spectra of Tc and Pc single crystals. The spectra
are shown at different pump-probe delay times for photoexcitation at
3.0 eV at room temperature for (a) Tc crystal in transmission and
(b) Pc crystal in
reflection. $\mbox{S}_{0}\rightarrow\mbox{S}_{1}$ and $\mbox{S}_{N}%
\rightarrow\mbox{S}_{1}$ ($\mbox{T}_{1}\rightarrow\mbox{T}_{N}$)
refers to electronic transitions in the singlet (triplet) manifold.
(See Fig. 1.)\vspace*{-10pt}}%
\label{Fig: 2}%
\end{figure}

High quality single crystals were grown in a flow of inert gas
\cite{Laudise1998}. The Pc crystals used were typically
3~$\times$~3~mm$^{2}$ and approximately 50 $\mu$m thick, while the
Tc crystals were larger. The Pc films were evaporated onto
10~$\times $~10~mm$^{2}$ MgO substrates, with a film thickness of
$\sim$150 nm, and $\sim$0.03\% molecular C$_{60}$-doping added to
one film. In these optical experiments we utilized a commercial
regeneratively amplified Ti:Al$_{2}$O$_{3}$ laser system operating
at 250 KHz producing nominally 10~$\mu$J, sub-50~fs pulses at
1.5~eV. The samples were excited at 3.0~eV (high above the
absorption band of $\sim$1.9~eV in Pc, and $\sim$2.4~eV in Tc), and
changes in reflectivity $\Delta R/R$ and transmissivity $\Delta T/T$
were measured over the range of probe photon energies from
$0.6-2.5$~eV using an optical parametric amplifier. We utilized an
optical chopper operating at 2 kHz together with lock-in detection
to measure relative changes with a sensitivity better than
10$^{-5}$. In all samples the signal displayed linear dependence on
excitation density in the range of fluences (F) used (in Pc
$\mbox{F}=10-200$~$\mu$J/cm$^{2}$, in Tc
$\mbox{F}=5-50$~$\mu$J/cm$^{2}$). The presented data was recorded at
$\sim$150~$\mu$J/cm$^{2}$ (Pc) and $\sim $30~$\mu$J/cm$^{2}$ (Tc).

The transient PI spectra from Tc and Pc crystals are shown in Fig.~2
at different times after photoexcitation. The most prominent feature
in both crystals is a long-lived ($\gg$~1~ns) PIA peak centered at
approximately $\sim$1.7~eV in Tc, and at $\sim$1.4~eV in Pc. The
long relaxation time suggests that the state being probed is the
triplet state T$_{1}$ ($\mbox{T}_{1}\rightarrow\mbox{T}_{N}$), i.e.
the PIA from the occupied T$_{1}$ level to a higher-lying unoccupied
excited state T$_{N}$ \cite{Tn}. Fig.~3 shows the corresponding
time-resolved dynamics for selected probe energies. In both
crystals, the sub-ps rise is followed by an initial decay process
after which a quasi-equilibrium is reached with a decay much longer
than 1 ns. This long-lived PIA is centered at $\sim$1.7~eV in Tc,
and at $\sim$1.4~eV in Pc \cite{dRRPc}. In Tc, the initial decay is
exponential and occurs on a 50-ps time scale. The recovery of the
PIA at low photon energies with a $\sim $50-ps exponential decay
(see the 0.64~eV data in Fig.~3(a)) can be associated with
$\mbox{S}_{1}\rightarrow\mbox{S}_{N}$ transitions and indicates a
large singlet exciton population which recombines to the ground
state. The data recorded at $\sim$1.7~eV reveals that some of the
population is transferred to the triplet manifold on a sub-ps time
scale (within the risetime). This is followed by a second process on
a 50-ps time scale, attributed to activated singlet fission,
$\mbox{S}_{1}\rightarrow2\mbox{T}_{1}$. The latter time scale
reflects the decay of the S$_{1}$ population. As Fig. 3(b) shows, in
Pc the dynamics are quite similar except for the fact that the
quasi-stationary state is established within a few ps. In Pc, the
internal conversion from $\mbox{S}_{N}$ to $\mbox{S}_{1}$ is
reflected by the 1.07~eV data with a $\sim$0.50-ps exponential
decay.

\begin{figure}[ptb]
\centering
\includegraphics[width=1.0\columnwidth]{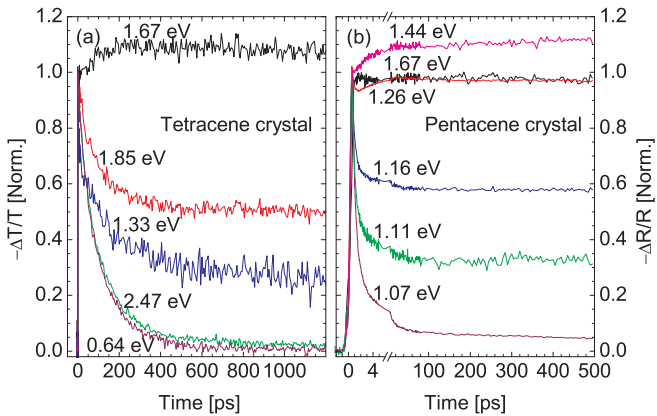}
\caption{Time-resolved PI transmission (reflection) dynamics of Tc
(Pc) single crystals at selected probe photon energies. The data is
normalized to the value recorded after the initial sub-ps rise.\vspace*{-10pt}}%
\label{Fig: 3}%
\end{figure}

The correspondence of the long-lived PIA to triplet production in Tc
(Pc) crystals is confirmed by probing the temperature dependence of
the $\mbox{T}_{1}\rightarrow\mbox{T}_{N}$ transition with a probe
energy of 1.67 eV (1.44~eV). Fig.~4 shows the temperature dependence
of the PIA signal at 200~ps after photoexcitation (when a
quasi-stationary state has been reached) which is proportional to
the population density of the $\mbox{T}_{1}$ level. While there is
no temperature dependence for Pc (which
is expected, since $E(\mbox{S}_{1})>2E(\mbox{T}_{1})$ and $E(\mbox{S}_{N}%
)>2E(\mbox{T}_{1})$), a pronounced temperature dependence is
observed for Tc. There is a substantial PIA signal present at the
lowest temperatures, indicative of a non-temperature dependent
triplet production process ($\mbox{S}_{N}\rightarrow2\mbox{T}_{1}$).
At temperatures above 200 K, the second contribution to the PIA
becomes important. It has a rise time of $\sim $50 ps (see inset to
Fig.~4), attributed to the thermally activated fission process,
$\mbox{S}_{1}\rightarrow2\mbox{T}_{1}$. The temperature dependence
follows the thermal activation law, $\Delta T/T(\mbox{T}, 200
\mbox{ps})=\mbox{A}+\mbox{B}\times\mbox{exp}[-\Delta\mbox{E}/k_{B}\mbox{T}]$.
Here A and B are constants, $k_{B}$ is the Boltzmann constant, T is
the temperature, and $\Delta\mbox{E}$ is the activation energy. We
find $\Delta\mbox{E}\sim70$ meV, in good agreement with previous
experimental values of the activation energy in Tc, $22-237$~meV
\cite{Tomki1971}.

The initial ultrafast dynamics observed on a sub-ps time scale
involves both
$\mbox{S}_{N}\rightarrow2\mbox{T}_{1}$ fission and $\mbox{S}_{N}%
\rightarrow\mbox{S}_{1}$ internal conversion in the singlet
manifold. In Tc, the time scale of the
$\mbox{S}_{N}\rightarrow2\mbox{T}_{1}$ fission is $\sim $0.3~ps at
300~K (decreasing to $\sim$0.25~ps at 5~K), while in Pc, it is
$\sim$0.7~ps and is temperature independent. Assuming an efficient
triplet production in Pc crystals (neglecting charge generation
\cite{Marciniak2007}), an initial singlet-exciton concentration of
$2.2\times10^{17}$~cm$^{-3}$ (calculated from the excitation
fluence) yields approximately $4.4\times 10^{17}$~cm$^{-3}$
triplets.

\begin{figure}[ptb]
\centering
\includegraphics[width=0.8\columnwidth]{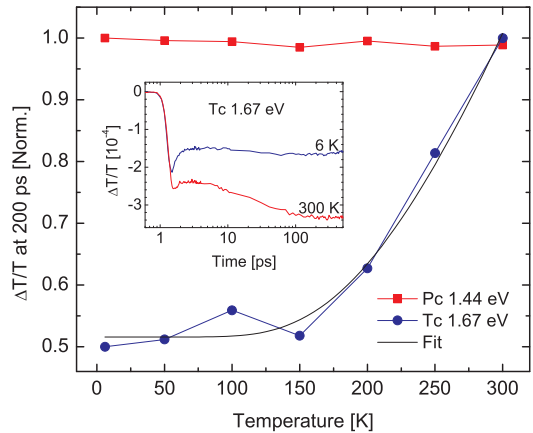}
\caption{Temperature dependence of the normalized PI signal in Tc
and Pc single crystals at 200 ps. The solid black line is a fit for
Tc: $\Delta T/T(\mbox{T}, 200 \mbox{ps})=0.52+7.77\times
\mbox{exp}[-\Delta\mbox{E}/k_{B}\mbox{T}]$, where
$\Delta\mbox{E}\sim70$ meV. The inset shows the time-resolved PI
changes in transmission for Tc probed at 1.67 eV at 6 K and 300 K.\vspace*{-10pt}}%
\label{Fig: 4}%
\end{figure}

The transient PI spectra for Pc films, shown in Fig.~5, resemble a
previously obtained spectrum published in ref. \cite{Marciniak2007},
and the steady state absorption spectrum (not shown) is identical to
the one given in ref. \cite{Ostro2005}. In contrast to Pc crystals,
the PI spectra for Pc films are dominated by sharp features at
higher energies, while the long-lived PIA centered at $\sim$1.4~eV
is substantially reduced. At room temperature we observe PI
bleaching peaks at $\sim$1.8~eV and $\sim$2.1~eV, and a PIA peak at
$\sim$2.0~eV. Their decay dynamics consists of two contributions; a
fast exponential $\sim$0.4 ps decay followed by a sub-ns decay. The
$\sim$1.8~eV and $\sim$2.1~eV peaks are assigned to
$\mbox{S}_{0}\rightarrow\mbox{S}_{1}$ and
$\mbox{S}_{0}\rightarrow\mbox{S}_{N}$ transitions, respectively,
while the PIA observed at $\sim$2.0~eV may be related to the
formation of excimer-like excitons, as argued in ref.
\cite{Marciniak2007}. At lower temperatures an additional PIA signal
appears at $\sim$1.9~eV (see inset to Fig. 5), which overshadows the
$\sim$2.0~eV peak observed at room temperature and persists well
into the ns regime. This strongly temperature-dependent PIA observed
at $\sim$1.9~eV is attributed to
$\mbox{S}_{0}^{+}\rightarrow\mbox{S}_{1}^{+}$
transitions, and is related to charge transfer dynamics. ($\mbox{S}_{0}%
^{+}\rightarrow\mbox{S}_{1}^{+}$ refers to an ionized state, where
an electron from the Pc molecule has been transferred to an electron
trap.) Importantly, the same dynamics and spectral features are
observed in the C$_{60}$-doped Pc film as shown in the inset to Fig.
5, except that the spectral features above $\sim$1.7~eV are enhanced
by $\sim$1.5~times, while the PIA at $\sim$1.4~eV is further
suppressed. Since C$_{60}$ is a known electron acceptor this
observation further lends support to the assignment of the
$\sim$1.9~eV absorption peak to charge transfer due to electron
traps intrinsic to Pc films.

\begin{figure}[ptb]
\centering
\includegraphics[width=0.75\columnwidth]{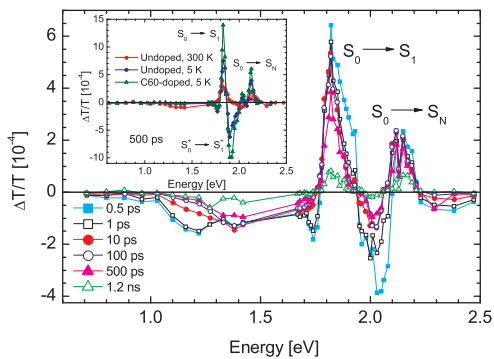}
\caption{Transient spectra of Pc thin film. $\Delta T/T$ is shown at
different pump-probe delay times following photoexcitation of 3.0 eV
at room temperature. The inset shows a comparison at 500 ps delay
and 5 K of the undoped Pc film and a C$_{60}$-doped Pc film.
$\mbox{S}_{0}^{+}\rightarrow\mbox{S}_{1}^{+}$ refers to an ionized
state at 1.9 eV due to electron transfer.\vspace*{-10pt}}%
\label{Fig: 5}%
\end{figure}

The long-lived $\mbox{T}_{1}\rightarrow\mbox{T}_{N}$ transition at
$\sim $1.4~eV, clearly prominent in the Pc crystal, is strongly
suppressed in the Pc film, Fig.~5, where only a minor feature with
sub-ns dynamics that vanishes at low temperatures is observed.
Considering the identical dynamics, temperature dependence and
enhancement of the spectral features in the C$_{60}$-doped Pc film
as compared to the undoped Pc film, we attribute the quenching of
the long-lived PIA at $\sim$1.4~eV (triplets) to ultrafast charge
transfer dynamics related to electron traps, as seen in the Pc
films.

In summary, we have shown that following optical excitation in Tc
and Pc single crystals triplets are produced via fission, not only
from the lowest singlet state,
$\mbox{S}_{1}\rightarrow2\mbox{T}_{1}$, but also from higher-lying
states, $\mbox{S}_{N}\rightarrow2\mbox{T}_{1}$. In Pc, fission is a
major relaxation channel, while in Tc there is strong radiative
fluorescence from $\mbox{S}_{1}$ which competes with
$\mbox{S}_{1}\rightarrow2\mbox{T}_{1}$ fission. At room temperature,
half of the triplets in Tc are produced immediately from
$\mbox{S}_{N}$ within $\sim$0.3~ps while the other half are produced
via thermally activated fission from $\mbox{S}_{1}$ within $\sim
$50~ps. This suggests that the internal conversion in Tc from
$\mbox{S}_{N}$ to $\mbox{S}_{1}$ is comparable to $\sim$0.3~ps,
which is the time scale for $\mbox{S}_{N}\rightarrow2\mbox{T}_{1}$
fission. Below $\sim$200~K the
$\mbox{S}_{1}\rightarrow2\mbox{T}_{1}$ fission is suppressed
consistent with a thermally activated process. In Pc, the triplet
production is temperature independent, and the majority of triplets
are produced within $\sim$0.7~ps via
$\mbox{S}_{N}\rightarrow2\mbox{T}_{1}$ fission, while
$\mbox{S}_{1}\rightarrow2\mbox{T}_{1}$ fission proceeds within a few
ps following internal conversion.

Considering the rise time dynamics in Pc and Tc crystals we conclude
that when
the condition for fission, which conserves spin, is fulfilled, $E(\mbox{S}_{1}%
)>2E(\mbox{T}_{1})$, triplet production may occur on a sub-ps time
scale. In thermally activated fission the time scale for triplet
production increases considerably ($>$100 ps for Tc). In comparison,
it is well-known that for intersystem crossing (which does not
conserve spin) the time scale may increase by orders of magnitude
\cite{ElectProc1999}. The ultrafast time scales demonstrated in Pc
and Tc may enable fast organic optical switching devices based on
triplet excitons.

Research and development of photoconversion-based devices such as
OLEDs and OPVs has largely involved polymeric and polycrystalline
materials in thin film morphologies and thus has not been concerned
with the influence of defects.  Here we have shown that defects can
cause triplet quenching and enhanced carrier generation.  In
particular, there is a large variation in triplet exciton dynamics
between Pc crystals and films, suggesting that a route to enhanced
triplet exciton lifetime is via the use of single crystals for
hosting the photoconversion process.  Thus, the present results
provide motivation for further research aimed at incorporating
crystalline material into the donor and acceptor regions of OPVs.

We acknowledge the support of the Laboratory Directed Research and
Development program at Los Alamos National Laboratory, the
Department of Energy (DOE) Center for Integrated Nanotechnologies,
and DOE grant number DE-FG02-04ER46118. We are grateful to Michael
Gr\"{a}tzel, Majed Chergui, Christoph Gadermaier and Thomas Dekorsy
for important comments.


\begin{references}
\bibitem {Tn}The position of the energy levels of the excited states in the
triplet manifold is still a matter of controversy, though it is
clear that the PIA originates from T$_1$.
\bibitem {dRRPc}In a thin sub-10~$\mu$m crystal we performed $\Delta
T/T$ measurements and found $\Delta T/T$ to follow $\Delta R/R$ in
sign and spectral shape. Since $\Delta T$+$\Delta R$+$\Delta A$=0,
it follows that $-\Delta R/R$ is proportional to $\Delta A/A$ (for
small changes).
\end{references}
\end{document}